# An Agile Method for E-Service Composition

Pouya Fatehi[1], Seyyed Mohsen Hashemi [2]

[1] Department of Computer Software , Science and Research Branch, Islamic Azad university
Tehran, Iran
*Pouya.Fatehi@gmail.com*

[2] Department of Computer Engineering, Science and Research Branch, Islamic Azad university
Tehran, Iran
*Hashemi@isrup.com*

**Abstract**
Nowadays, application of Service Oriented Architecture is increasing rapidly; especially since introduction of distributed electronic services on the web. SOA software has a modular manner and works as a collaboration of independent software components. As a result, e-service approach is sufficient for software with independent components, each of which may be developed by a different company. Such software components and their cooperation form a composite service. Agile methodologies are the best candidate for developing small software components. Composite services and its building blocks are small pieces of software, making agile methodology a perfect fit for their development.
In this paper, we introduce an agile method for service composition, inspired by agile patterns and practices. Therefore, across the agile manifesto, we can develop low cost, high quality composite services quickly using this method.

## 1. Introduction

rends to inter enterprise communication and creation of on demand-services, made e-services more popular technology. Another reason of e-service popularity is Software as a Service (SaaS) model which is a rapidly growing model of software licensing. *"SaaS is seen as a possible replacement to traditional software where the buyer obtains a perpetual license and installs and maintains all necessary hardware, software and other technical infrastructure. Under SaaS, the software publisher (seller) runs and maintains all necessary hardware and software and buyers obtain access using the Internet".*[1]
A *Service* is a discoverable software resource which has an advertised service description. The service description is available in a repository, called a registry that can be searched by a potential consumer of the service. Given the service description, a service consumer can bind to and use the service. In real world e-services are represented by web service which has different types (grid services, distributed services, etc).
Making advanced features and goals, companies uses a combination of several services to develop composite service. The composability feature of services leads to low-cost development of high quality services. *"The process of integrating existing Web services to achieve higher-level business tasks that cannot be fulfilled by any individual service alone is often referred to as Web service composition"*[2]

Based on SOA modeling on IBM [3] service composition is 4[th] step of SOA modeling. Service Identification, Service Specification, Service Realization are earlier steps and last one is Service Implementation. Whereas services are small software components, different software methodologies could be used for developing a composed service.
RUP and similar guides, like the PMBOK, Software Engineering Institute's (SEI's) Capability Maturity Model Integrated (CMMI), or the UK's IT Infrastructure Library (ITIL) standards impose unnecessary process overhead for smaller projects. In contrast agile methodologies allow for fast and tight increments or phases; cut down overhead; and ensure a close relationship between developer and customer. Comparing agile methodologies with other methodologies, demonstrates that agile methodologies is more useful for developing services. So in this paper we are trying to represent a new agile methodology for services; especially composed Services.
To explore which agile methods are the most effective for composite service development, we first investigate if any combinative use of agile methods (or, at least, employing more than one agile method) can have an effect on the four outcomes. Based on research that has been done in [4], using two agile methods is better in productivity rather than using only one method, and there appeared to be no significant further advantage in increasing the number of methods used beyond two. Also [4] expressed that in terms of both quality and productivity, there was a significant difference between the eXtreme Programming/Scrum combination and all the other pairs of methods. However there was no significant difference in either cost or satisfaction. This clearly tells us that the eXtreme Programming/Scrum combination is a good pairing of methods to adopt.
This result can be seen to make some sense in that, eXtreme Programming (XP) is very much oriented towards technology based practices and programmer activity. In contrast, Scrum is more focused on agile project management aspects[15] .In addition, Scrum is explicitly intended as a wrapper around other engineering approaches. Therefore XP and Scrum can be seen to be complementary from a practical point of view, supporting the claims made by Mar and Schwaber (2002)[16].
Accordingly we use a combination of these pair methods to bring agility in our service composition model.Since changing development process can not implemented in a single shot, we defined two implementation levels. By defining separate levels in development process, preparation of implementation process would be simple like what

CMM[1] done. Also IBM defined a capability maturity model for services (SIMM) [5]. This model consists of 7 steps and companies should implement dynamic service composition to achieve 7th step. First level of AM4SC also is useful for describing our model strategy.

## 2. Agile Model for Service Composition (AM4SC)

AM4SC described by 4 activities which derived from Scrum methodology and 6 practices which comes from XP method. Mainly activities are management process and practices increases agility and development quality. The 4 activities are described as below:

**1- Planning process:** In this process functionality domain described and with TDD rules an abstract model for composed service would be designed. As a result expected inputs and outputs are represented. Obviously according to agile manifesto, proposed model would not be a complete model but it is started with a simple model and will be finalized with iteration and incremental in later evolves.

**2-Arrangement process:** After planning process, features list that should be included in next iteration prepared. This process will be done at a periodical session and as a output of services and features that should be included in next iteration (something like backlog in scrum) reported. Some of the services listed above, should be refactor to fulfill the composed service rules.

**3-Modeling process:** As in the previous steps, expected input, expected output and requirement list determined, modeling could be done with any available service modeling tools; In the simplest case, by using pen and paper. The test process can be started from here. Across to the input, output and designed model in first step, tests can be introduced before implementing composed service. Each service will be considered as a block and their outputs are determined in each specific input.

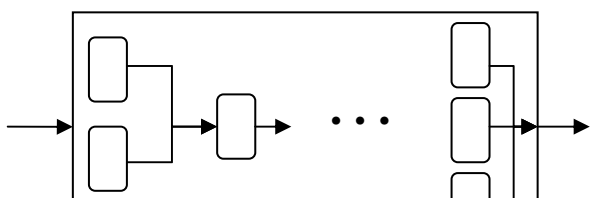
*Fig1 Composite service modeling*

**4- Implementation Process:** At this step the platform of independent model for composite service designed in previous step mapped to platform specific model. The result is a composite service that covers requested functionality. In the last, this result delivered in this iteration to customers.

This life cycle will be continued by receiving customer feedback and their new requirements and composite service incrementally completed in each iteration.

Practices bring more agility to AM4SC. 6 practices are based on XP practices and make our model in a line with agile manifesto as bellow:

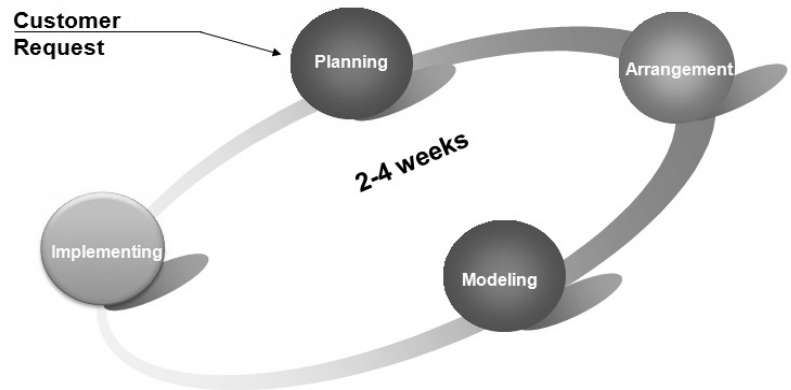
*Fig 2 AM4SC Activities*

**Collaborative programming:** Developed services may implement in different platforms and with different technologies by several development teams. Although service description is available through UDDI and with Web Service Description Language (WSDL), but developers of services can share their experiences for modeling and implementing composite service in low cost at easiest way.

**Iterative and Incremental development:** Requirement should be categorized. So main features developed in first iteration then with refactoring other features would be inserted to composite service incrementally.

**Online customers:** composite service customers also called service requester. Their feedbacks will be used to increase quality and acceptance of developed service. Also distributing beta versions in web, delivers more feedbacks.

**Continues integration:** In each iteration, whole service model should be integrated to implement composite service, this makes our model more reliable.

**Reusability through design:** This is a fundamental agile practice that involves actively ensuring that a piece of functionality is implemented in a single place. We use these techniques to make sure that services which used in composite service, implement functionality once. Also same techniques could be used to brings more functionality and cover more requirements with less services. Design reviews are used to remove duplicate logic across existing services. In order to increase the potential of reusing the services, ensure that they do not have channel, technology platform or transport specific logic which is coupled with the service's domain logic.[5]

---
[1] Capability Maturity Model

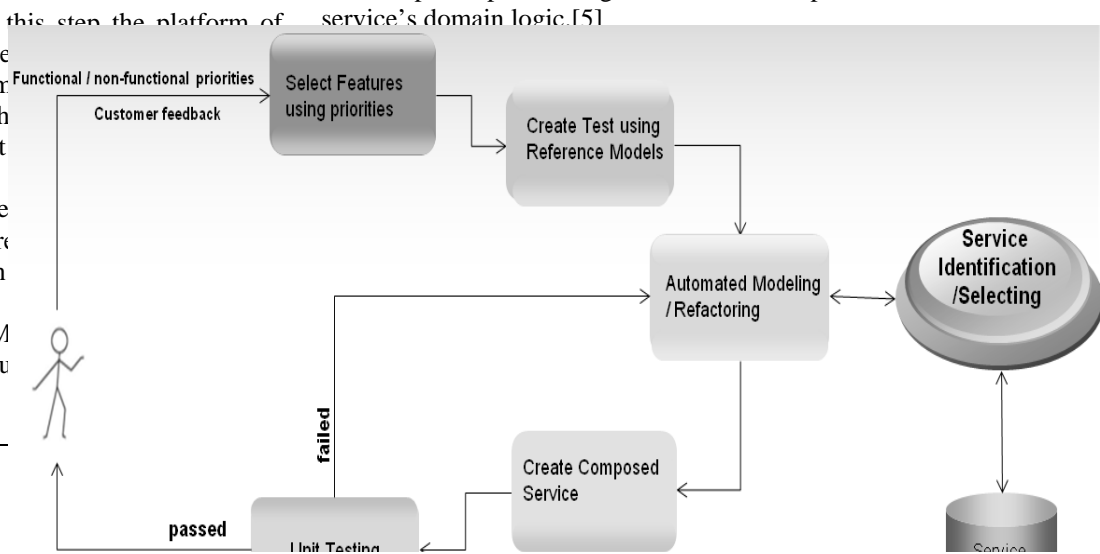
*Fig3- collaboration of AM4SC agents*

Today, service developers use SaaS for distributing their service in the web. To use this method through service development, the AM4SC method should use a dynamic method for developing composed services. For reaching this goal, we defined an automated model for AM4SC that not only have a same agility feature of first level but also brings capability of developing composed service in run time and on demand.

The execution of AM4SC starts with pair element of request and request priorities defines by service requester. After modeling and developing compose service, despite of unit test acceptance, a new version of service delivers to the customer. In AM4SC second level, six agents collaborate together to achieve model goals as bellow:

At first, an agent called feature selector, select features by pairs that service requester sent as an input. Additionally based on meet in the middle architecture, customer's requests should be categorized across the developer's policies and their priorities which provided by requester. Therefore in this agent, customer's request or feedback priorities with company policy variables would be used for selecting features that will be develop at the next iteration to decrease risk of development. Output of this agent, also is useful tool for accessing progress of development process.

Second agent which named unit test maker, searches among reference models for defined problem domain that described by service requester requirements. This agent acts as a TDD practice to increase quality of developed service. Unit tests, generated by this agent would be used as artifacts to describe requirements of the problem.

Then automated refactor/modeler agent, creates requests for binding required service based on selected features and artifacts that defined by unit test maker agent. The request should first translated to WSDL format then requests would be sent to service broker agent that search UDDI registry for requested services, delivers binding information of services to automated modeler agent. Then automated modeler use an automated modeling method for model composed service using bind information. The model simply can be converted to BPL or BPEL format and then creation process of service will be started y next agent.

Another task that would be done by this agent is model refactoring when unit tests fails or in next iteration of service development, cycles to add more features in composed service.

Refactoring is one of main principle of DRY practice that is responsible for locate opportunities for eliminating duplication among services. Therefore refactoring agent also can be used to increase reusability of composed service.

At the next step, an agent named Service Creator binds services and creates needed relationship designed in service model between input and outputs of services that described by BPEL.

Tester agent, Tests composed service with created unit tests and surveys the results. If Tests pass, the composed service delivers to service requester, otherwise if one or more tests failed, model with failed unit tests delivered to automated modeler for refactoring and changing model to satisfy all unit tests. As for unit tests are artifacts that describes features, if unit tests pass we can ensure that all planned features have been covered.

At the end of iteration, after designed unit tests passed, an integrated composed service would be delivered to service requester. Customer's feedback and recommendations could be used to improve productivity and usability of composed service in next iteration of developing composed service.

*Table 1 Comparison of AM4SC with other composition methods*

| Composition Types | Methods | 1- Service composition support | 2- Meet in the middle support | 3- Participation of stake holders and customers | 4- Repetitive production | 5- Categorizing requirements | 6- Managing creation process | 7- Fast release | 8- Optimal usage of available resource | 9- Reusability | 10- Refactoring | 11- Continues integration | 12- Simple Design | 13- flexibility | 14- Periodical release | 15- Dynamic Composition | 16 - Availability of Techniques | 17- Modeling capability | 18- Quality Assurance |
|---|---|---|---|---|---|---|---|---|---|---|---|---|---|---|---|---|---|---|---|
| SOA Methodology | SOMA[6] | ✓ | ✓ | | | | | | | | | | | | | | | | |
| | OASIS Methodology[7] | | | ✓(2) | ✓(1) | ✓ | ✓ | | | | | | | | | | | | |
| | Microsoft Motion Methodology[8] | | ✓ | ✓ | ✓ | | | ✓ | ✓ | | | | | | | | | | |
| Agile methods in SOA | MSMUA(5)[5] | ✓ | ✓ | | ✓ | ✓(4) | | ✓ | | ✓ | ✓ | ✓(3) | ✓ | ✓ | | | | | |
| | SRI-DM[9] | ✓ | | | ✓ | | | | | | ✓ | | ✓ | ✓ | ✓ | | | | |
| Service Composition Methods | SCDC(6)[10] | ✓ | | | | | | | | | | | | | | ✓ | | | |
| | Multi Agent model[11] | ✓ | | | | | | | | | | | | | | ✓ | ✓ | | |
| | Model Based[12] | ✓ | | | | | | | | | | | | | | ✓ | ✓ | | |
| | Ant colony based model[13] | ✓ | | | | | | | | | | | | | | ✓ | ✓ | | |
| Agile Methodologies | XP Scrum FDD AMDD DSDM ,…[14] | | ✓ | ✓ | ✓ | | ✓ | ✓ | ✓ | | ✓ | ✓ | ✓ | ✓ | ✓ | | ✓ | | ✓ |
| Offered Method | AM4SC | ✓ | ✓ | ✓ | ✓ | ✓ | ✓ | ✓ | ✓ | ✓ | ✓ | ✓ | ✓ | ✓ | ✓ | ✓ | ✓ | ✓ | ✓ |

---

[1] In first level once a year and in second level after fourth iteration
[2] In collaborative working practice
[3] Done-Done practice
[4] With user stories and their updates
[5] Modern SOA Methodology and SOA Adoption Using Agile Practices
[6] A Service Compatibility Model to Support Dynamic Cooperation of Cross-Enterprise Services

## 4. Comparison of AM4SC with other Service Composition Methods

In this part we are considering to compare AM4SC to other service composition methods from agility and main service composition features perspective. Some of these methods (like SOMA, OASIS, etc) are not specially designed for service composition but covers service composition as a part of SOA modeling.

Table1 represents these comparisons. Some features (4, 6, 11, 9, 10, 12, and 19) cover productivity and reliability of composed service while some others are for decreasing risk of production. In next section we are describing that how these features are supported by AM4SC.

1- **Service composition:** In our comparison we take look at other SOA methodology to sure that we are supporting their features. Although some of these methods don't include composition process, but supports all other service development features.
2- **Meet in the middle methodology**: features that would be generated at next iteration should be selected by features selector agent based on user priorities and company policies to decrease risks.
3- **Participation of customers**: Customers feedbacks and priorities cause to low risk reliable software. AM4SC infrastructure is on demand development method that composes services to satisfy customer requirements. Customers also, send his request priorities to service development process that would be used for selecting next iteration features which should be developed.
4- **Repetitive production:** by periodical and repetitive production that described above, service stability and quality would be guaranteed.
5- **Categorizing requirement**: development would be started from simple model and completed in several iterations. So we will sure about main featured that should be developed and then consider about other features.
6- **Managing creation process**: As for AM4SC is derived from scrum methodology process will be managed in a good manner.
7, 12-**Fast release/simple design**: Development of composite service started with some base features, and final product would be completed incrementally. This kind of development caused to fast product delivery.
8- **Optimal usage of available resource:** One of the agile manifesto rules is optimal usage of human resources for developing software in low cost. These 6 agents are responsible for 5 activities that described above which are minimum required activities that should be used to cover both scrum and XP methodology.
10- **Reusability/Refactoring**: Refractor agent is responsible for redesigning of composite service model without logical change in service features as described before. As composed service, may use as a component of another composed service; so reusability of service through design should be considered. Based on [5], user stories and refactoring are two main practices that would be caused to increase reusability of service.
11- **Continues integration**: In each iteration, whole composite service integrated for delivering a fresh version to customer. Several periodical iteration, increase quality of service and reliability by decreasing side effects of new inserted features.
13- **Flexibility**: New feature should be added based on customer's feedback and requests in each iteration. So composed service is so flexible that could take a look at all features based described priorities.
14- **Periodical release:** in each period we have a fresh version of composed service and consumer can choose best one that fit his/her requirements.
15- **Dynamic composition support :** using agents that automatically perform service composition tasks, enables on demand composition and as a result, composed service can be described in run time.
16- **Availability of techniques:** Since all agents duties and practices derived from agile methods, these techniques are available and tested before.
17- **Modeling capability:** Modeling that will be done by automated modeler, increase readability for both customer and developers.
18- **Quality assurance:** Test driven development that would be done by unit test maker and unit tester agent is best practice that satisfies quality assurance of development

## 5. Conclusions

In this paper an agile method based XP and Scrum methodology used to brings more agility for service composition process. So service production process become faster and more flexible, additionally all other agile production features will be achieved.

### References
[1] G Vidyanand Choudhary, "Software as a Service: Implications for Investment in Software Development"

**Pouya Fatehi** is software designer in [Iran Rayaneh](#) Company and is responsible for service architecture of BPM project which names FilerPlus for 5 years.
He has a Masters degree in Software Engineering from Science and research Beranch of Islamic Azad university of Tehran. He is specialist in software as service paradigm and has a several research in the field of agile methodologies and SOA.

**Seyyed Mohsen Hashemi** received the M.S. degree in Computer Science from Amirkabir University of Technology (Tehran Polytechnic University) in 2003, and the PhD degree in Computer Science from the Azad University in 2009. Moreover, he is currently a faculty member at Science and Research Branch, Azad University, Tehran. His current research interests include Software Intensive Systems, E-X systems (E-Commerce, E-Government, E-Business, and so on), Global Village Services, Grid Computing, IBM SSME, Business Modeling, Agile Enterprise Architecting through ISRUP, and Globalization Governance through IT/IS Services..